# The failure of DFT-based computations for a stepped-substrate-supported correlated Co wire

Nader Zaki, Hyowon Park, Richard M. Osgood, Andrew J. Millis, Chris A. Marianetti
Columbia University, New York, NY 10027, USA

Density functional theory (DFT) has been immensely successful in its ability to predict physical properties, and, in particular, structures of condensed matter systems. Here, however, we show that DFT qualitatively fails to predict the dimerized structural phase for a monatomic Co wire that is self-assembled on a vicinal, i.e. stepped, Cu(111) substrate. To elucidate the nature of this failure, we compute the energetics of a Co chain on a Cu surface, step, notch, and embedded in bulk. The results demonstrate that increasing Co coordination extinguishes the dimerization, indicating that the failure of DFT for Co on the Cu step arises from excessive hybridization, which both weakens the ferromagnetic correlations that drive the dimerization and increases the bonding that opposes dimerization. Additionally, we show that including local interactions via DFT+$U$ or DFT+DMFT does not restore the dimerization for the step-substrate supported wire, though the Co wire does dimerize in DFT+DMFT for the isolated vacuum case. This system can serve as a benchmark for future electronic structure methods.

# I. INTRODUCTION

Density functional theory (DFT) computations have been immensely successful in predicting physical properties of condensed matter systems throughout the periodic table. However, DFT computations in all known implementations are found to qualitatively break down in certain strongly correlated electron systems [1], sometimes predicting a metal when the system is experimentally observed to be an insulator. Nonetheless, even in these rare scenarios when DFT computations qualitatively fail to describe the electronic ground state, the crystal structure is typically predicted correctly. A recent example of a qualitative failure to predict structure was in the rare earth nickelates, where the experimentally observed Ni-O bond disproportionation [2] is not found to be an energy minimum in DFT. However, incorporation of local intra-$d$-orbital interactions via the so-called DFT+$U$ method [3] improves agreement substantially [4], while treating the same interactions via the density functional plus dynamical mean field method (DFT+DMFT) produces a phase diagram in quantitative agreement with experiment [4]. Here we present a more flagrant failure. We show the experimental observation of a dimerized Co chain on a Cu surface [5] eludes DFT, DFT+$U$, and DFT+DMFT computations. In previous work, we demonstrated that the dimerization was driven by ferromagnetic correlations [5]. We expand upon this work and demonstrate how and why DFT, DFT+$U$, and DFT+DMFT fail. This physical example can serve as a testbed for future beyond-DFT total energy methods.

# II. METHODS

DFT calculations were performed using the Vienna Ab-Initio Simulation Package (VASP) with a plane-wave basis and the projector augmented wave (PAW) method [6, 7].

Details of the calculations for the Co wire in vacuum case are detailed elsewhere [5]. For all calculations, the spin polarized generalized gradient approximation (GGA) functional was used. Calculations were performed for different substrate geometries as shown in Fig. 1: Co wire on Cu(111), Co wire on Cu(332), Co wire embedded in Cu(111), and Co wire embedded in Cu bulk. For the surface slab calculations, a periodic system with a vacuum spacer of at least 10Å was used. We have compared 2-layer Co wire/Cu(332) – i.e. 6 step rows, 8-layer Co wire/Cu(332), and 2-layer Co wire/Cu(775) – i.e. 7 step rows – calculations, and observe similar energy/phase behavior. Hence, we are confident that the 2-layer Cu surface substrate results presented here will not differ significantly from that utilizing more substrate layers or even/odd number of step rows. For the surface slab calculations, at least 3×1×12 *k*-points were used to sample the surface Brillouin zone, where the *z*-direction was *parallel* to the Co wire, the *x*-direction was *perpendicular* to the Co wire and the surface normal, and the *y*-direction was along the surface normal. We found that increasing the number of *k*-points beyond this configuration yielded a negligible energy difference of about 20meV/slab (~0.8 meV/atom) or less. Since dimerization is driven by Kohn-Sham eigenvalues that lie close to the Fermi level, Brillouin-zone integrations were performed via the tetrahedron smearing method with Blöchl corrections [8]. A plane-wave energy cutoff of 400eV was used for all surface slab calculations. While this work focused on infinite length chains, calculations were also performed for finite length chains and are discussed in the Appendix.

In previous work [5], we demonstrated that ferromagnetic correlations were the driving force for dimerization of the Co chain. DFT predicts a strong preference for ferromagnetism, which results in spin-minority $d_{xz}/d_{yz}$ derived bands that are half-filled and hence susceptible to a Peierls distortion. Modeling the wire system using an expansion of the total energy in terms of spin cluster functions is illuminating in our understanding of the role of frontier Co $d$-orbitals as well as the elastic term to dimerization [5]. Any lattice observable (in our case, the energy) which is a function of discrete site variables, can be expressed as a power series expansion in the correlation functions of site variables [9]. The contribution of the frontier Co $d$-electrons to the dimerization can be accurately represented using nearest neighbor and next-nearest neighbor correlation functions of the spins on the Co sites [5], as defined in the following equation:

$$H = E_0 + \sum_i J_1\, s_{2i} \cdot s_{2i+1} + J_2\, s_{2i+1} \cdot s_{2i+2} + J_3\, (s_{2i} \cdot s_{2i+2} + s_{2i+1} \cdot s_{2i+3}) \tag{1}$$

where $E_0$ is the non-magnetic or elastic energy contribution, $s_i$ is $\pm 1$, the spin orientation on site $i$, and $J_1/J_2$ are the nearest-neighbor magnetic pair interactions, while $J_3$ is the next-nearest-neighbor magnetic pair interaction. An illustration of this model is shown in Fig. 2. For the Co/Cu system, DFT results show the spin moments reside on the Co sites, and hence this model works equally well for a Co wire on substrate as it does in vacuum. This is evidenced by the small fit errors of less than 1meV/atom for all substrate cases investigated here.

### III. RESULTS AND DISCUSSION

#### A. DFT

We start with the DFT predicted optimal Co wire position on a vicinal substrate, which is obtained using a full relaxation of the Co atoms while holding the Cu substrate atoms fixed. This calculation was performed for an 8-layer Cu(332) slab as well as a 2-layer Cu(332) slab, (step width of ~12Å). Identical results for the Co site positions were obtained, confirming that a 2 layer substrate can be used. For the 8-layer case, an additional calculation was performed in which the top 6 Cu atom layers were allowed to also fully relax, while the bottom 2-layers were fixed. For the fixed substrate case, the relaxation results show the Co wire sinking vertically into the Cu terrace by ~0.12Å (5.4%) and horizontally into the step edge by ~0.05Å (2.4%) relative to the starting point of the perfect Cu lattice. For the relaxed substrate case, the relaxation results show the Co wire sinking horizontally into the step edge by ~0.08Å (3.6%); the vertical sink was a negligible 0.02Å (<1%). In all cases, however, the results do not show a structural distortion of the Co wire (the substrate relaxed case showed a relatively trivial distortion of 0.03Å, or ~1%, which differed from the non-distorted case by a mere 0.3meV). Instead, the Co atoms are equally spaced at 2.56Å, the spacing of the underlying ideal Cu. In experiment, the Co wires dimerize with a short bond length of 2.0Å (78% of mean spacing). Thus DFT qualitatively fails to describe the experimental result.

DFT calculations for the unrealistic case of a Co wire in vacuum do predict dimerization, of the magnitude experimentally observed for substrate supported Co wires [5]. This result provides us with a means of studying why the method fails for a more realistic slab calculation. We have studied Co wires with varying degrees of coordination to a supporting substrate; these configurations in order of increasing coordination number were (0) a Co wire in vacuum [5], (1) a Co wire on top of a Cu(111) slab, (2) a Co wire at the step edge of a Cu(332) slab (i.e. a stepped slab), (3) a terrace embedded slab, in which the Co wire makes up one of the rows in a Cu(111) slab, and (4) an embedded matrix of Co wires in bulk Cu, where each Co wire makes up the 6th row of the slab. These calculations are periodic in the plane of the slab. For cases (1), (3) and (4), the Co wires are ~13.3Å apart, while for case (2), they are ~12.0Å apart. A diagram of the slabs for these different configurations is shown in Fig. 1. Unless otherwise noted, the Co atoms are located in the ideal site positions of the corresponding substrate Cu atoms.

For each model geometry, we computed the energy as a function of the short bond length in a dimerized wire. Results are shown in Fig. 3. We find that the Co wire weakly dimerizes for case (1), insignificantly for case (2), and not at all for case (3). Hence, under DFT, the degree of dimerization decreases with increasing coordination with the substrate. To obtain further insight we performed calculations for different magnetic configurations and fit the results to Eq. 1. The corresponding cluster fits are plotted in Fig. 4; we have also plotted the magnitude of the derivative of the magnetic and elastic energy terms with respect to bond length.. As was previously found for the case of a Co wire in vacuum [5], the net magnetic energy term, shown in blue, drives the dimerization, while the non-magnetic term (the elastic term), shown in red, discourages it. The plots make it clear that for the case of a substrate-supported wire, the driving magnetic term decreases with increasing coordination number, reflecting a change in the Co wire electronic structure due to hybridization with the substrate, while the inhibiting elastic term increases with increasing coordination number, reflecting increased pinning of the Co positions to the Cu position; a detailed view of this trend is provided in Fig. 5. This analysis of the weakening of the magnetic energy term and the strengthening of the elastic energy term indicates that DFT is over predicting the binding (i.e. hybridization) of the wire to the substrate.

### B. Extensions of DFT

Having demonstrated the failure of DFT calculations, a logical step to correct for this is to use the DFT+$U$ method [3], wherein a local intra-$d$-orbital interactions, not fully captured by DFT, are treated in a Hartree approximation. DFT+$U$ has become a popular extension of DFT that has been shown to enhance the magnetic and orbital moments as well as correct structural failures of DFT and is now widely used, though the fact that it is a Hartree-Fock approximation means that it can, at times, overemphasize polarization. We have investigated the Co wire system using DFT+$U$ and have found that it too fails to provide correct qualitative predictions. In fact, even for the case of a Co wire in vacuum, dimerization is not predicted under DFT+$U$. To understand why dimerization is not favored, we refer to the Kohn-Sham eigenvalue/band diagram for the undimerized Co wire in vacuum case, Fig. 6, calculated using $U = 4eV$ and $J = 1eV$. The DFT+$U$ bands are seen to be quite different from the DFT bands, in agreement with previous calculations [10]. While spin-polarized DFT predicted nearly ½-filled $d_{xz}, d_{yz}$ spin-minority orbitals, which drove the dimerization via a Peierls-like mechansim, DFT+$U$ predicts that these two orbitals are completely filled; the $d_{z^2}$, which also assisted in driving the dimerization, is rendered completely empty. DFT+$U$ does produce a nearly ½-filled $s$ orbital, but this does not strongly promote dimerization due to its large bandwidth. Thus, the problem with DFT+$U$ is an unphysical orbital occupancy.

Since it is known that DMFT can capture both the localized and itinerant aspects of electron dynamics, we adopted the DFT+DMFT method [4], for the total energy calculation of a Co chain in vacuum. Fig. 7 shows the DMFT result for the dimerization energy, as a function of bond length and calculated for two spin cases: first, by enforcing paramagnetic (PM) spin symmetry and second, by allowing for ferromagnetic (FM) spin symmetry. FM DMFT calculations show that the dimerization is indeed favored, with energy gain comparable to (though smaller than) that found in the plain (non-orbitally polarized, see Supplementary Material of [5]) GGA result. The PM DMFT calculation dimerizes, but not as strongly as the FM DMFT result. . The dimerization energy is reduced to 0.14eV and the length of the short bond is increased to ~2.2Å (86% of the undistorted value). Therefore, we conclude that it is essential to use calculation methods which allow for ferromagnetic intersite correlations, but do not erroneously break orbital symmetry, for a proper treatment of dimerization in this Co wire system.

We have also performed total energy calculations for the step-supported case using the DFT+DMFT method. The DMFT calculations were performed imposing both the PM spin symmetry and the FM spin symmetry, as in the isolated chain case. As shown in Fig. 7, we found that the DMFT calculation with FM spin symmetry does not favor the dimerization; nor does the PM case dimerize. The PM ground state energy was found to be slightly higher than that of the FM case. Similar to the DFT result, DFT+DMFT essentially fails due to excessive hybridization with the substrate. While it is true that on-site correlations do reduce hybridization with the substrate via many-body renormalizations, they also reduce the inter-site ferromagnetic correlations which are responsible for the dimerization. While it is true that on-site correlations do reduce hybridization with the substrate via many-body renormalizations, they do not change the energetics significantly. Therefore, we conclude that single-site DFT+DMFT cannot capture the experimental dimerization of the Co chain on the Cu substrate.

## IV. CONCLUSIONS

While DFT is generally *qualitatively* correct for predicting structural parameters, we have shown that DFT computations fail qualitatively to predict the dimerized structural phase for a monatomic Co wire that is self-assembled on a vicinal Cu(111) substrate. This failure is due to DFT's over-prediction of hybridization of the Co wire with the underlying Cu substrate. We used a cluster expansion to demonstrate that this over-hybridization leads to weakening of the magnetic coupling along the wire, which is necessary for dimerization, while increasing the stiffness of the wire due to strengthening of the non-magnetic elastic term. We demonstrate that the DFT+$U$ method also fails due to erroneous orbital polarizations induced by this approximation. Furthermore, we also investigated the dynamical mean field (DMFT) method of correcting the DFT calculations, considering both paramagnetic and ferromagnetic solutions. We found that while the DFT+DMFT method removes the unphysical orbital ordering predicted by DFT+$U$, and confirms the association between dimerization and ferromagnetic nearest-neighbor correlations, DFT+DMFT is always less favorable to dimerization than the pure DFT calculations, and in particular predicts that the step-substrate supported Co chain does not dimerize. A Co chain on a Cu step should be viewed as a new test case for beyond-DFT total energy methods such as DFT combined with cluster-extended dynamical mean-field theory.

## V. ACKNOWLEDGEMENTS

This work was supported by U.S. Department of Energy, Office of Basic Energy Sciences, Division of Materials Sciences and Engineering under Award Contract No. DE-FG 02-04-ER-46157. AJM was supported by DOE-ER-046169. This research utilized resources at the New York Center for Computational Sciences at Stony Brook University/Brookhaven National Laboratory which is supported by the U.S. Department of Energy under Contract No. DE-AC02-98CH10886 and by the State of New York.

## VI. APPENDIX

### A. Finite length monatomic wires

While the calculations in this work have focused on monatomic wires of infinite length, we note that slab supported calculations have been performed for the case of step-substrate supported finite length chains [11]. Pick *et al* found that for finite length chains, calculated up to 7 atoms, a CDW dimerization instability did not occur; instead, a varying amount of anisotropic strain was found. We have performed structural relaxation calculations for chains up to 10 atoms in length, using a 2 layer Cu step slab geometry and restricting relaxation to only the Co chain atoms, and have obtained similar results as Pick *et al*. The closest resemblance to a dimerized Co pair was found for the case of a 2-atom Co chain (i.e. a true dimer), which had a bond length contraction that was approximately 60% of the experimentally measured dimerization contraction. For chains longer than two atoms, only the pair of atoms at the end of the chain showed significant collinear contraction; the contraction due to this end effect was approximately 44% of the experimentally measured dimerization contraction.

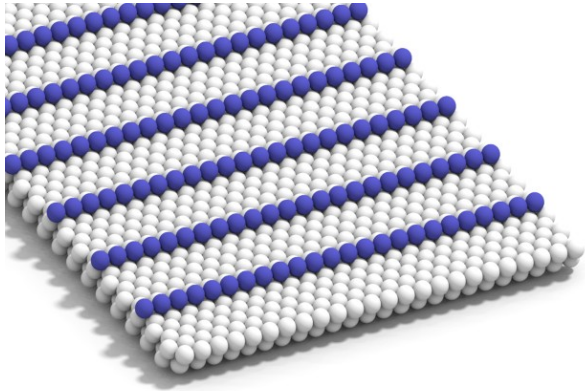 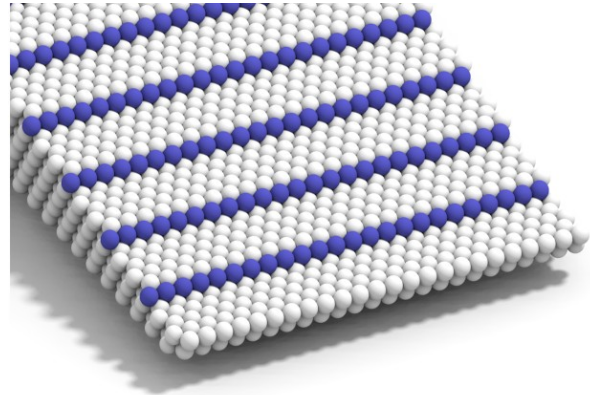

**Cu(111)**  **Cu(332)**

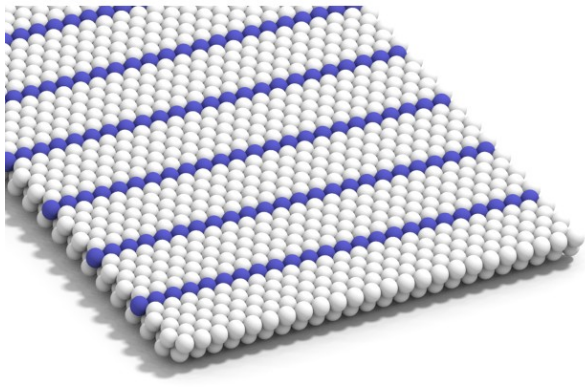 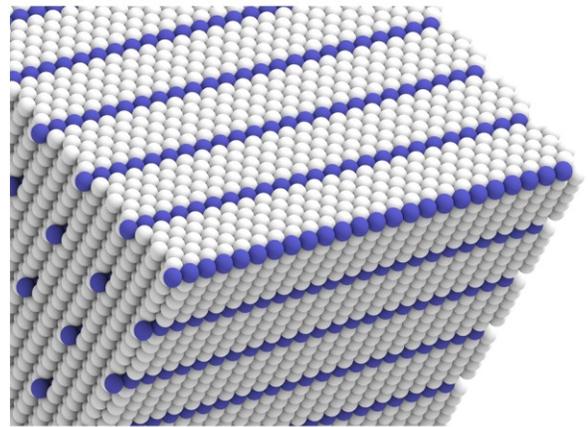

**Cu(111) wedged**  **Cu bulk**

**Figure 1** Different substrate geometries on which slab calculations for a Co 1-D wire have been calculated. Cu atoms are colored white while Co atoms are blue. For the Cu(111), Cu(111)-wedged, and the Cu bulk cases, the Co wires make up every 6$^{th}$ row, which gives a wire-to-wire separation of ~13.3Å. For the vicinal substrate case, Cu(332), the step terrace width is ~6 atom rows, and the wire-to-wire distance is ~12.0Å. In this work, these slab geometries are denoted as cases (1)-(4), respectively.

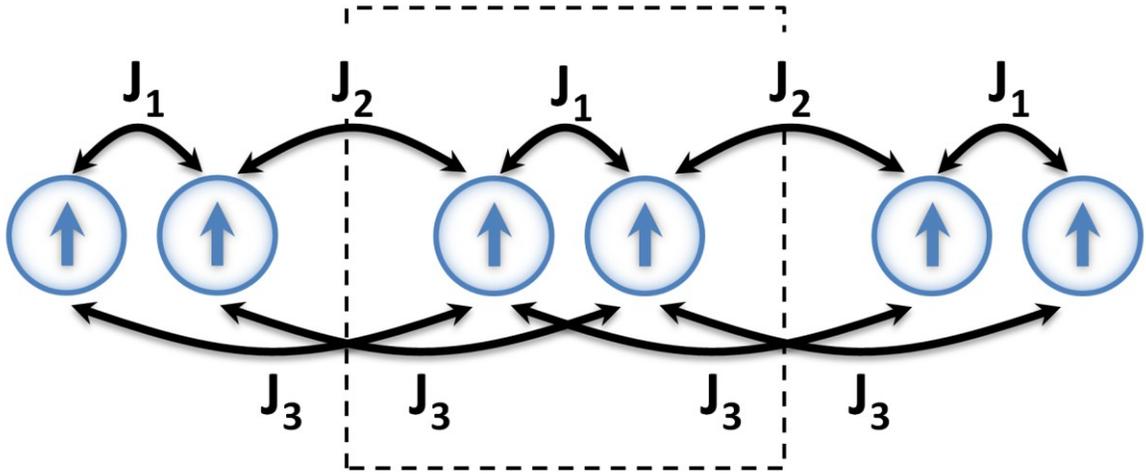

**Figure 2** An illustration of the cluster expansion model utilized in this work. $J_1$/$J_2$ are the nearest neighbor magnetic pair interactions, while $J_3$ is the next-nearest neighbor magnetic pair interaction. The dashed rectangle denotes one unit cell.

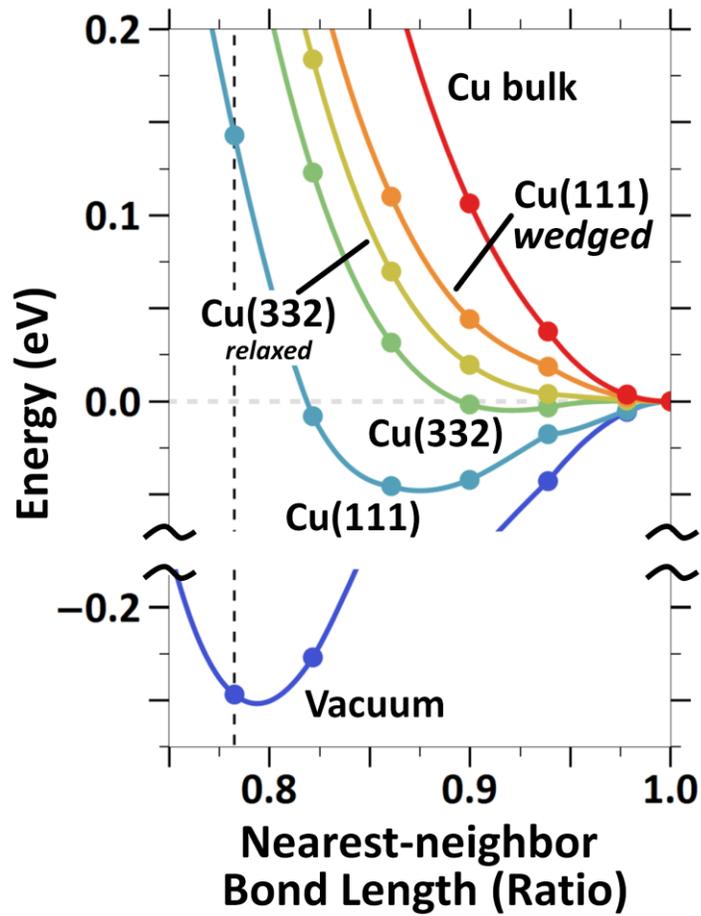

**Figure 3** Energy phase diagram for a Co wire in different slab configurations. The horizontal axis is the ratio of the short bond length of the Co wire with respect to the non-distorted bond length (2.56Å). The relaxed Cu(332) curve corresponds to the case where the Co wire is positioned at the optimally relaxed position with respect to the Cu(332) substrate. The dashed vertical line denotes the experimentally measured short bond length for a Co wire on vicinal Cu(111).

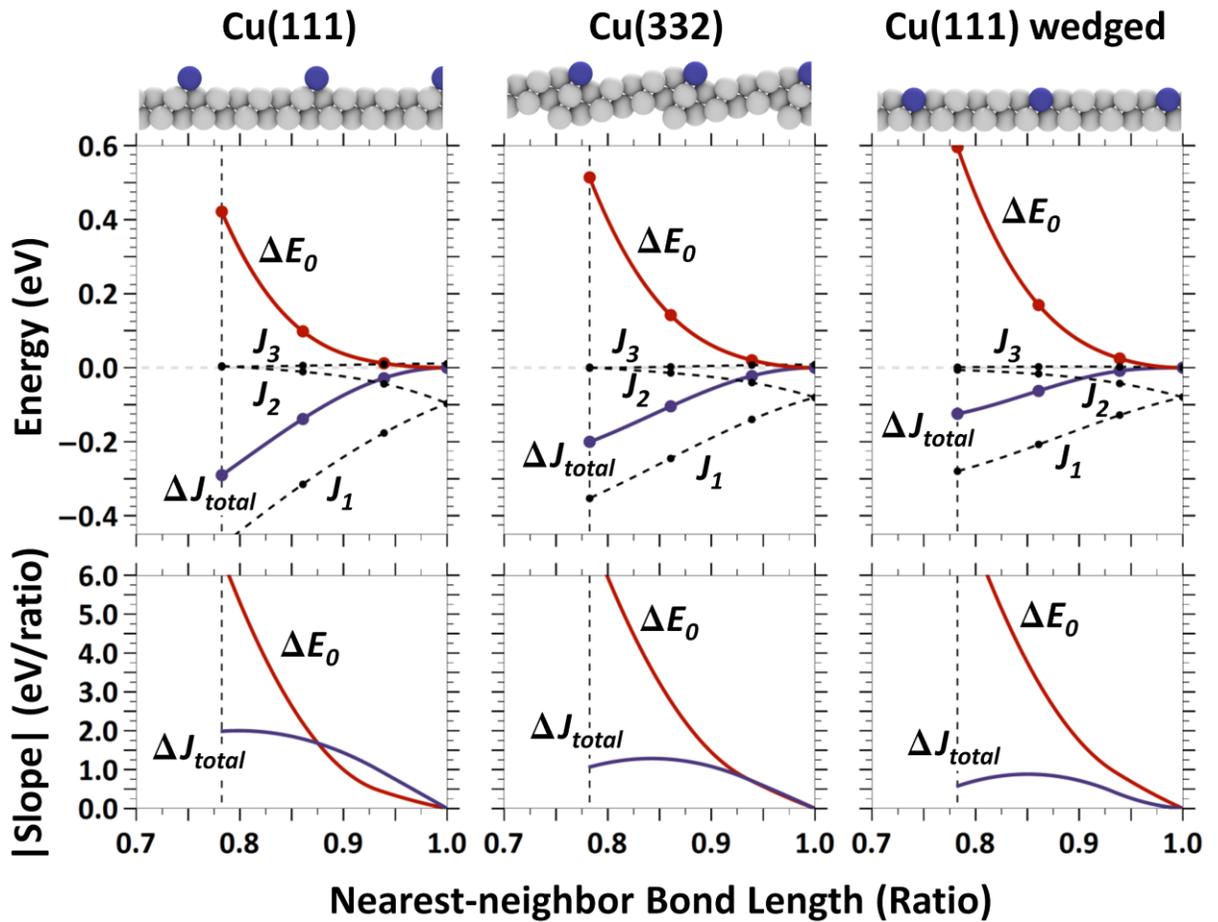

**Figure 4** Cluster expansion parameter fits for different Co wire slab configurations. The horizontal axis is the ratio of the short bond length of the Co wire with respect to the non-distorted bond length (2.56Å). The bottom row of plots correspond to the magnitude of the slope of the magnetic and non-magnetic energy contributions shown in the above row plots as the blue and red curves, respectively; this is obtained from the 1st derivative of the interpolation of the points. The dashed vertical line denotes the experimentally measured short bond length for a Co wire on vicinal Cu(111).

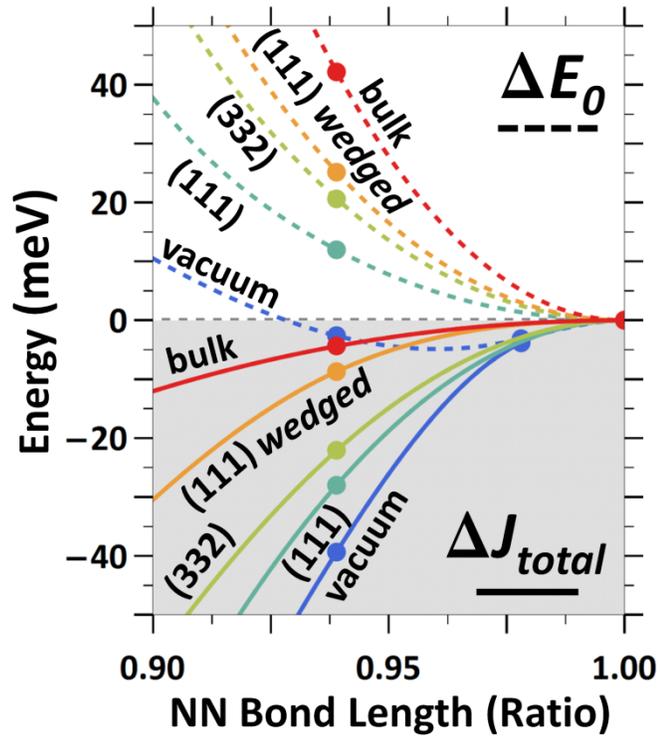

**Figure 5** An expanded plot of the cluster expansion parameter fits for different Co wire slab configurations, establishing the trend with increasing coordination. The horizontal axis is the ratio of the short bond length of the Co wire with respect to the non-distorted bond length (2.56Å). The solid colored lines correspond to the total magnetic energy contribution; the dashed colored lines denote the non-magnetic energy contribution.

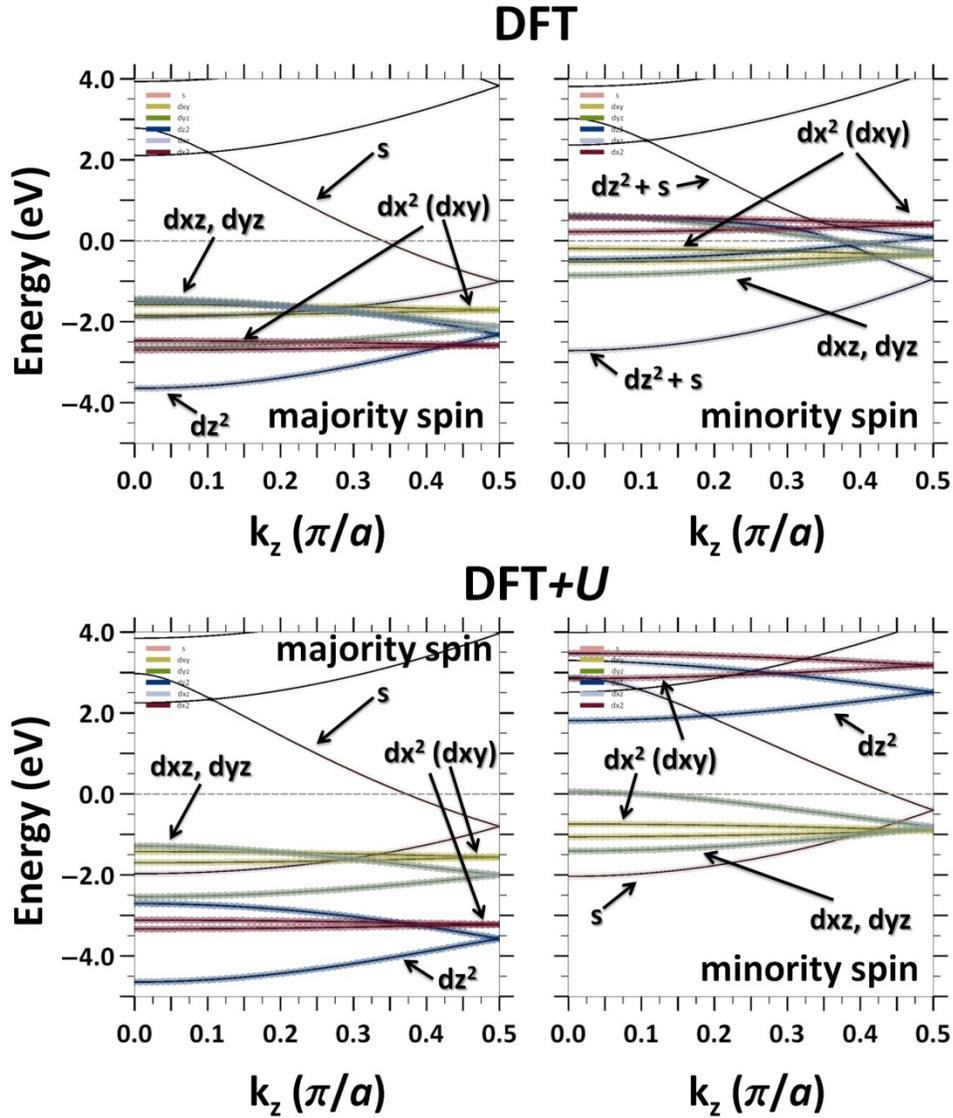

**Figure 6** Orbital projected Kohn-Sham eigenvalue band diagram for a Co wire in vacuum (2 atoms per unit cell). The Co atom spacing is the non-distorted bond length (2.56Å). The top row plot corresponds to the spin polarized GGA DFT calculation while the bottom row corresponds to the DFT+$U$ version with $U = 4eV$ and $J = 1eV$.

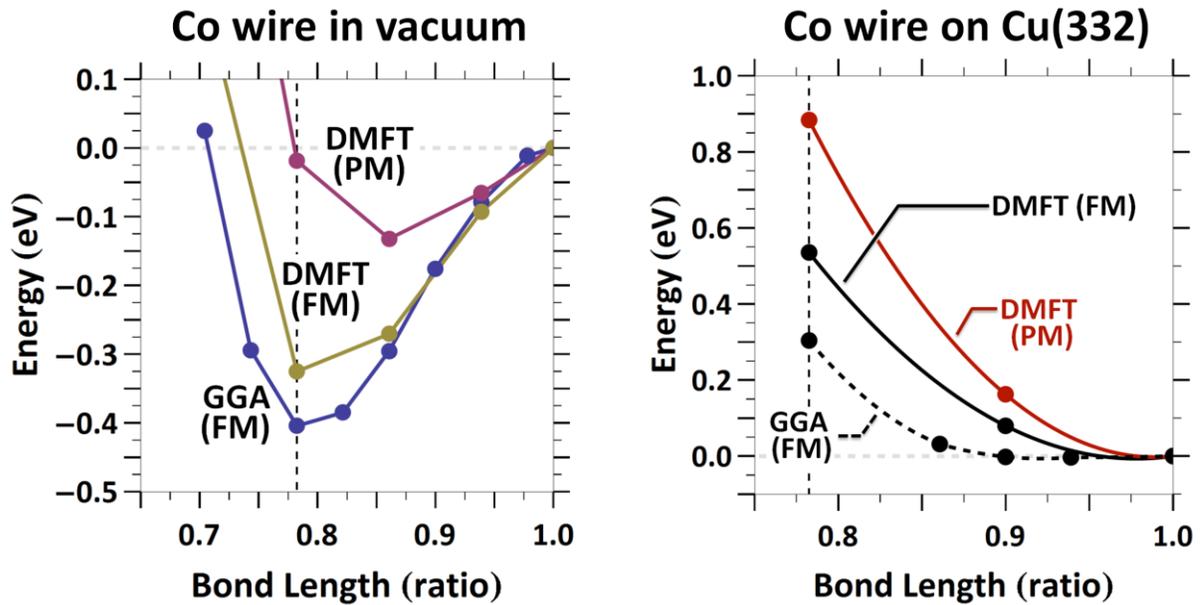

**Figure 7** Energy phase diagram for the isolated Co wire and step-substrate, Cu(332), supported Co wire under different amounts of distortion using DMFT. For ease of comparison, corresponding DFT results are plotted here as well. For DMFT, two spin configurations, paramagnetic and ferromagnetic, are shown, while for DFT, the ferromagnetic results are shown (for the case of the isolated wire, the non-orbitally polarized case is used). The horizontal axis is the ratio of the short bond length of the Co wire with respect to the non-distorted bond length (2.56Å). The dashed vertical line denotes the experimentally measured short bond length for a Co wire on vicinal Cu(111).